\documentstyle[prb,manuscript,aps,epsfig]{revtex}

\newcommand{\qeff}{q_{\mbox{\scriptsize{eff}}}}
\newcommand{\barqeff}{\bar{q}_{\mbox{\scriptsize{eff}}}}

 \tightenlines

\begin{document}
\draft
\author{Hans J\"org Limbach and Christian Holm }
\address{
  Max-Planck-Institut f\"ur Polymerforschung,
  Ackermannweg 10, D-55128 Mainz, Germany }
\title{End-effects of strongly charged polyelectrolytes - a molecular dynamics
  study}
\date{\today}
\maketitle
\begin{abstract}
  We investigate end-effects in the ion distribution around strongly charged,
  flexible polyelectrolytes with a quenched charge distribution by molecular
  dynamics simulations of dilute polyelectrolyte solutions. We take the
  counterions explicitly into account and calculate the full Coulomb
  interaction via an Ewald summation method.  We find that the free
  counterions of the solution are distributed in such a way that a fraction of
  the chain charges is effectively neutralized.  This in turn leads to an
  effective charge distribution which is similar to those found for weakly
  charged titrating polyelectrolytes that have an annealed charge
  distribution.  The delicate interplay between the electrostatic
  interactions, the chain conformation and the counterion distribution is
  studied in detail as a function of different system parameters such as the
  chain length $N_m$, the charge fraction $f$, the charged particle density $\rho$,
  the ionic strength and the solvent quality.  Comparisons are made with
  predictions from a scaling theory.
\end{abstract}
\vspace{2ex} 
\pacs{PACS: 61.25.Hq Macromolecular and polymer solutions;
  polymer melts; swelling, 36.20.Ey Conformation (statistics and dynamics),
  87.15.Aa Theory and modeling; computer simulation }

\narrowtext

%
%
\section{Introduction}
%
%

``Polyelectrolytes are polymers bearing ionizable groups, which, in polar
solvents, can dissociate into charged polymer chains (macroions) and small
counterions'' \cite{barrat96a}. The combination of macromolecular properties
and long-range electrostatic interactions results in an impressive variety of
phenomena which makes these systems interesting from a fundamental as well as
from a technological point of view.

A thorough understanding of polyelectrolytes has become increasingly important
in biochemistry and molecular biology. This is due to the fact that virtually
all proteins, as well as the DNA, are polyelectrolytes.

Unfortunately, the theoretical understanding of polyelectrolytes is less
developed than the understanding of the properties of neutral polymers. The
presence of long-range interactions renders the application of renormalization
group techniques and scaling ideas much more difficult than in the neutral
case. This is due to the fact that many new length scales appear that are not
well separated and therefore can influence each other in a complicated fashion.

The degrees of freedom of the counterions contribute largely to the entropy.
Their equilibrium distribution is strongly influenced by the immobile charges
on the macroion. However, this distribution is strongly coupled to the
conformation of the macroion itself, resulting in a complex interplay of chain
conformation and ion distribution. In the following we will present a
systematic investigation of the spatial counterion distribution around
strongly charged flexible polyelectrolytes by means of computer simulations.

In this work we will treat the case of quenched strongly charged
polyelectrolytes in both good and poor solvent for the backbone which possess
a fixed charge distribution along the polymer backbone that is usually fixed
through the initial chemistry. This is the case, for example, for the often
studied polystyrene sulfonate \cite{essafi95a,essafi96a}.  Strong means here
that the average charge separation on the backbone is on the order of the
Bjerrum length so that we are in the regime were some of the counterions are
in close proximity to the macroion, a phenomenon that is usually called
counterion condensation \cite{manning69a,oosawa71a}.

Through molecular dynamics simulations (MD) we will demonstrate explicitly
that the counterion distribution around a quenched strongly charged
polyelectrolyte shows the appearance of an ``end-effect'', namely that the
distribution of counterions around the ends is significantly different from
that around the inner part of the chain.  A similar situation has been
analyzed theoretically in a recent work by Castelnovo et al.
\cite{castelnovo00a} for annealed weakly charged polyelectrolytes. Annealed
means that the degree of ionization of the macroion can depend on the pH of
the solution, so that the ionization sites of the backbone can move along the
chain.  In their paper they show that the electrostatic field of the
polyelectrolyte alters the degree of ionization along the backbone, thus
leading to an ``end-effect'' in the charge distribution.  This effect has
already been found by numerical simulations \cite{ullner96a,berghold97a} for
polyelectrolyte chains interacting via a Debye-H\"uckel potential. It could be
of some relevance for systems where end-effects are known to be important,
like adsorption on charged surfaces \cite{zhulina95a}, or self-assembly of
weakly charged linear micelles \cite{schoot97a}.

We will show that a totally flexible, strongly charged polyelectrolyte system
with a quenched charge distribution shows qualitatively the same behavior. We
argue that this follows from the fact that the positions of the counterions
are ``annealed''. Thus the distribution of all charges along the chain (fixed
monomer charges and mobile ions) appears to be annealed.  The result is again
an ``end-effect'' that originates physically in the electrostatic field of the
polyelectrolyte. We thus demonstrate that the qualitative picture of the
charge distribution which emerges out of the scaling theory is even correct
for strongly charged polyelectrolytes in a regime where the blob picture
ceases to be meaningful, because for the highly charged case the electrostatic
blobs are of the order of one monomer.

The paper is organized as follows: First we describe our model and the
simulation technique. Then we present our approach to analyze the end-effects.
Afterwards we show our results for systems of flexible polyelectrolyte chains
in good solvent with explicit counterions in the dilute concentration regime,
and investigate the influence of the chain length, the charge parameter, the
density and the ionic strength on the charge inhomogeneity along the chain
contour length.  The last section treats the case of a poor solvent chain, and
we end with some conclusions.

%
%
\section{Model and simulation technique}
%
%
Our model of a polyelectrolyte solution consists of $N_p$ flexible
bead-spring-chains with $N_m$ monomers, $N_c$ counterions and in some cases
$N_s$ pairs of salt ions which are located in a simulation box of length $L$
with periodic boundary conditions (3D torus).  A fraction $f$ of the $N_m$
monomers is monovalently charged ($v=1$).  The number $N_c$ of monovalent
counterions ($v=-1$) is then chosen such that the overall system is
electrically neutral.  If we have $N_q = N_m f$ charges on a polymer chain,
then the number of counterions is given by $N_c = N_p N_q$, and the total
number of charged particles in the system is $N_{tq}= 2 (N_c+N_s)$, giving
rise to a charged particle density $\rho = \frac{N_{tq}}{L^3}$.

The interaction between monomers is described via a standard Lennard-Jones
potential of the form:
\begin{equation}
U_{LJ}(r) = \left\{ \begin{array}{c@{\quad : \quad}l}
4\cdot \epsilon_{LJ} [(\frac{\sigma}{r})^{12} - (\frac{\sigma}{r})^{6}
- c(R_c)]
& \mbox{for } r \le R_{c}  \\[0.1cm]
\hspace*{0.95cm} 0 & \mbox{for }  r > R_{c} \\
\end{array} \right.
\label{ljpotphob}
\end{equation}

The function $c(R_c)$ is chosen as $c(R_c) = (\frac{\sigma}{R_c})^{12} -
(\frac{\sigma}{R_c})^{6}$ to give a potential value of zero at the cutoff.
For the good solvent case all chain monomers interact only via the repulsive
part of the Lennard-Jones potential, hence $R_c = 2^{1/6}\sigma$. For the poor
solvent case we set $R_c = 2.5 \sigma$, which gives the chain monomers a short
range attraction. The discontinuity at $R_c$ is small compared to the applied
random forces and causes therefore no problems for the stability of the
simulation.  In this case the depth of the potential minimum can be tuned by
$\epsilon_{LJ}$.  For both cases we assume that all ions do not have any short
range attractive parts in the Lennard-Jones interaction, as it is reasonable
for alkali metals.

The chain monomers are in addition connected along the chain by the FENE
(finite extendible nonlinear elastic) bond potential,
\begin{equation}
U_{\rm FENE}(r) = -\frac{1}{2} k R_0^2 \ln \left(1 - \frac{r^2}{R_0^2}\right)
\label{fenepot}
\end{equation}
with spring constant $k=7 \frac{k_B T}{\sigma^2}$ and finite extension
$R_0=2\sigma$.  All charged particles interact in addition via full Coulombic
interaction
\begin{equation}
E_c (r_{ij}) = \lambda_B k_B T \frac{ v_i v_j }{ r_{ij} },
\end{equation}
where $v_i$ is the valence of the $i^{th}$ charged particle in units of the
elementary charge $e$ and $\lambda_B=\frac{e^2}{4 \pi \epsilon_0 \epsilon_r
  k_B T}$ is the Bjerrum length characterizing the strength of the
electrostatic interaction.

The electrostatic energy of the box is calculated with the P3M
(Particle-particle particle-mesh) algorithm which is based on the Ewald
summation method. Details of this method can be found in Deserno and Holm
\cite{deserno98a,deserno98b}.

In our molecular dynamics (MD) simulations we had no explicit solvent
molecules. However, we implicitly take the polarizability of the medium into
account through an effective relative permeability $\epsilon_r$, which we take
to be that of water at room temperature, $\epsilon_r = 80$.

The (MD) method employed in the present work is similar to the one used in
Ref. \cite{grest86a}. To simulate a constant temperature ensemble, the
particles are coupled to a heat bath. The motion of the $i^{th}$ particle is
given by the Langevin equation: \(
m\frac{d^{2}}{dt^{2}}\overrightarrow{r_{i}}=-\overrightarrow{\nabla }V_{tot}(
\{\overrightarrow{r_{j}}\} )-m\Gamma
\frac{d}{dt}\overrightarrow{r_{i}}+\overrightarrow{f_{i}}(t) \), where
\textit{m} (chosen as unity) is the mass of the particles, \textit{\( V_{tot}
  \)} is the total potential force made up of the above described
Lennard-Jones, FENE and Coulomb terms, which are all pairwise additive, \(
\Gamma \) denotes the friction coefficient, and \( \overrightarrow{f_{i}} \)
is a random force. The two last quantities are linked by the
dissipation-fluctuation theorem \( <\overrightarrow{f_{i}}(t)\cdot
\overrightarrow{f_{j}}(t')>=6m\Gamma k_{B}T\delta _{ij}\delta (t-t^{'}) \).
We used a damping constant $\Gamma = \tau^{-1}$, with time step $0.015\tau$ at
constant temperature $k_B T=1\epsilon$.

The number of MD steps was chosen such that the typical observables like the
end-to-end distance $R_e= \sqrt{\langle \vec R_e^{2} \rangle}$ or the radius
of gyration $R_g$ had sufficiently relaxed, which happened usually after
500\,000 up to 2\,000\,000 MD steps.  We normally performed between  
5\,000\,000 and 10\,000\,000 MD steps to take measurements.
Some basic parameters and
observables of the investigated systems are summarized in Table
\ref{basisdata}. The charge parameter $\xi$, often referred to as Manning
parameter, is defined as the number of unit charges along the chain contour
per Bjerrum length.  Since the chains are flexible we have also given an
effective charge parameter $\xi_{R_e}$ which gives the number of unit charges
per Bjerrum length when the whole chain is mapped on a rod of length $R_e$.
For more information about chain extension in such systems see e.g. Stevens
and Kremer \cite{stevens95a}.

%
%
\section{Data Analysis of the effective charge}
%
%
Our strongly charged polyelectrolyte chains have a charge parameter $\xi$ of
the order of unity, hence some fraction of the counterions are located closely
to the chain (they would be Manning condensed, if the chain was infinitely
long \cite{manning69a}). The main idea of defining an effective charge is that
the chain charges can be effectively neutralized by the close proximity of an
oppositely charged counterion, which is basically the concept of charge
renormalization \cite{alexander84a}. Our definition of the effective charge
which we will explain in this paragraph is meant as a straight forward
practical approach.  This is due to the difficulty of defining this quantity
in a rigorous theoretical framework.  Thus we do not fix the effective charge
by fitting the asymptotic distribution to a Debye-H\"uckel distribution of a
renormalized spherical charge distribution as in Ref.\cite{alexander84a}, nor
do we actually require the ions to obey some Manning condensation criterion
since that is only defined for infinite rod-like polyelectrolytes
\cite{manning69a,deserno00a}.

To investigate the proximity between a free ion $i$ and a polyelectrolyte
chain we introduce $d_i(j)$, the distance between this ion and the charged
monomer $j$ that is closest in space.  ${\cal C}_j$ is then the set of all
ions $i$ which are closest to monomer $j$.  To enhance statistics we sometimes
average over neighboring monomers.

\begin{figure}[htb] \begin{center}
    \epsfxsize=10cm \epsfbox{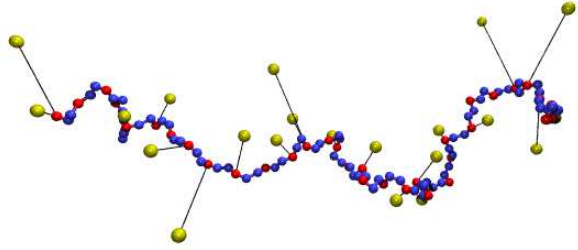}
  \end{center}
  \caption{Snapshot of a polyelectrolyte chain of system 1. The Figure shows
    the geometrical situation in the vicinity of the chain. The method of
    measuring the distance of free ions to the chain and their assignment to
    individual monomers is illustrated by the connections between the ions and
    their closest monomer}
  \label{geometry}
\end{figure}

A snapshot of a polyelectrolyte and the counterions in its vicinity
illustrating the definition of their smallest distance $d_i(j)$ is shown in
Fig.  \ref{geometry}. For a reasonably stretched charged polymer the ions are
located inside an approximately cylindrical space and are assigned to every
charge along the contour length of the polyelectrolytes.  The average ion
charge $n_q(r,j)$ located at a distance $r$ from the charged monomer j is then
given by

\begin{equation}
n_q(r,j) = \left\langle \sum_{i \in {\cal C}_j}  v_i \delta (r - d_i (j))  \right\rangle
\label{n_q(r)}
\end{equation}
Here $\delta(x)$ is the Dirac delta function.  The brackets indicate
the canonical average which is taken over all chains and configurations.

By integrating $n_q(r,j)$ from zero to $r$ we obtain
\begin{equation}
P(r,j) = \int\limits_{0}^{r} \mbox{d}r^{\prime} \quad
n_q(r^{\prime},j)
\label{P(r)}
\end{equation}
which is the local ion charge contained in the interval $[0,r]$.  We are now
able to define the local effective charge
\begin{equation}
\qeff(r,j) = (1+P(r,j))
\label{q_eff(r_c)}
\end{equation}
as a function of monomer $j$ and distance $r$.

The concept of an effective charge applied to the entire polyelectrolyte
instead of each charged monomer yields the effective charge as a function of
radius alone.  The average ion charge $n_q(r)$ located at a distance $r$ from
the polyelectrolyte is then calculated via equation (\ref{n_q(r)}) by taking
the sum over all charged monomers.  Via equation (\ref{P(r)}) and
(\ref{q_eff(r_c)}) we can define also the integrated counterion charge $P(r)$
and the effective charge $\qeff(r)$ for the polyelectrolyte.  Later we will
need those quantities normalized to the chain charge $N_q$,

\begin{equation}
\barqeff(r) = \frac{1}{N_q} \qeff(r)
\qquad \mbox{and} \qquad
\bar{P}(r) = \frac{1}{N_q} P(r)
\label{ave_q_eff(r_c)}
\end{equation}

Of course there is a difficulty in choosing for $r$ an appropriate cut-off
radius $r_c$. For our purposes it is not necessary to require that the ions
are condensed in the physical sense of Manning, which is rigorously applicable
anyhow only for the case of infinitely long rods (see for example the
discussion in Ref. \cite{deserno00a}).  It is well known that the concept of
an effective charge for macromolecules is difficult to define both theoretical
and experimental and that it depends on the properties which one considers
\cite{belloni84a,schmitz93a,viovy00a}.  Since we are interested in the
deviation of the effective charge along the chain backbone from its mean value
we found that the qualitative result is rather independent from the way the
effective charge is defined. Thus we have chosen a simple operational
definition and investigate in the next paragraph how different cut-off radii
influence the end-effects.

\begin{figure}[htb]
  \begin{center}   
    \input{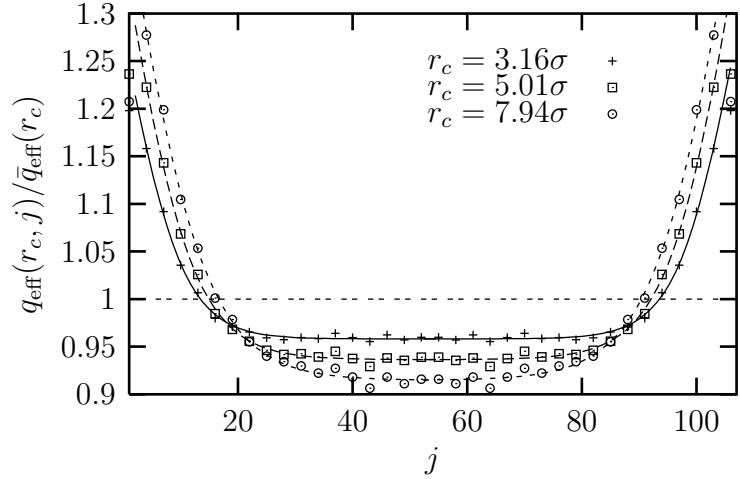}
  \end{center}
  \caption{Inhomogeneity of the effective charge along the contour length:
    Qualitative independence from the cutoff radius $r_c$.  System 1. As
    throughout the paper the fitted lines are thought as a guide to the eye.}
  \label{qeff_rc}
\end{figure}

In Fig. {\ref{qeff_rc}} the relative effective charge
$\qeff(r_c,j)/\barqeff(r_c)$ is shown for three different values of $r_c =
3.16 \sigma , 5.01 \sigma , 7.94 \sigma $ (see System 1 in Table \ref{basisdata}).
One observes that the qualitative form is independent from the exact value of
$r_c$, although $\barqeff(r_c)$ changes slowly from $0.72$ over $0.64$ to
$0.57$ with increasing $r_c$, simply because more counterions are accounted
for.  More pronounced is the decrease of the relative effective charge at the
very ends with increasing $r_c$ due to the increasing geometrical artifacts,
namely the pole-caps become larger.  Another good reason to choose a small
$r_c$ is the need to look at the chain as a rodlike object, as is explained
more clearly in the next paragraph, and the fact that $r_c$ should be smaller
than the Debye screening length $\lambda_D^{-2} = 4 \pi \ell_B (c_{ci} + 2
c_s)$ which is especially important at high densities (high counterion
concentrations $c_{ci}$) and high salt concentrations $c_s$. In the case
considered in Fig.~\ref{qeff_rc} the screening length assumes the value
$\lambda_D = 23.0\sigma$. The advantages of a larger cut-off are that more
counterions are found within $r_c$, which enhances the statistics.  However,
as can be seen from the scattering of the data points, the statistical error
is quite constant for the three values of $r_c$.  For most of our data
analysis we have decided to fix $r_c = 5.01 \sigma $, thus dropping the $r_c$
dependence from our notation.

In the following discussion of the results we characterize the curves for the
relative effective charges $\qeff(r_c,j)/\barqeff(r_c)$ with help of some
basic quantities.  The first, which we call amplitude, is the maximal
difference of $\qeff(r_c,j)/\barqeff(r_c)$ between an end and the middle of
the chain. Here we want to remark that the maximal value of $\qeff(r_c,j)$ is
often not achieved at the outermost monomers due to the above mentioned
geometrical artifacts.  The amplitude contains information about the strength
of the end-effect.  The second is the penetration depth of the end-effect,
which is defined through the occurrence of a plateau of
$\qeff(r_c,j)/\barqeff(r_c)$ in the middle of the chain. For example, in Fig.
\ref{qeff_rc}, this plateau extends approximately from $j=25$ to $j=80$ which
corresponds to a penetration depth of $25$ monomers along the contour. Using
$R_e$ this contour length can be rescaled to a length in space. In our example
the contour length of $25$ monomers corresponds to a real extension of
$10.8\sigma$.  One should keep in mind, however, that the absolute values are
weakly varying functions of the chosen cut-off $r_c$.

%
%
\section{Results}
%
%
In this section we will discuss our main results for the ion distribution and
its inhomogeneities around flexible polyelectrolytes under various conditions.
First we will compare our results with analytical theory.  Then we will have a
closer look on the influence of various parameters such as the chain length,
the charge parameter, the density, the salt concentration and the solvent
quality.  We will also investigate conformational inhomogeneities which can be
seen in the bond energy distribution.

\subsection{Connection to the Cell Model \label{cellmodel}}

Even though we want to focus on the inhomogeneities of the counterion
distribution around polyelectrolytes we first look at the integrated ion
charge $P(r)$ itself. This quantity can be exactly computed within mean-field
Poisson-Boltzmann theory for the cell model ($CM$) of an infinitely long
charged rod\cite{fuoss51,alfrey51}, and the solution is characterized by the
three parameters, namely the rod of radius $r_0$, the charge parameter
$\xi_{CM}$, and the enclosing cylindrical cell of radius $R_C$.  A recent
discussion of this model and comparisons with computer simulations can be
found, for example, in the work of M. Deserno \cite{deserno00a}.

\begin{figure}[htb] 
  \begin{center}   
    \input{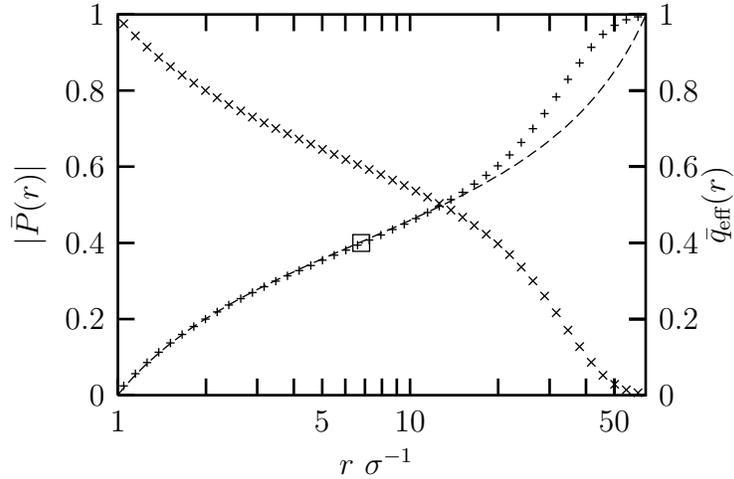}
  \end{center}
  \caption{Ion charge distribution $|\bar{P}(r)|$ (+) and effective charge
    $\barqeff(r)$ ($\times$) as a function of the radius $r$. The dotted line
    is a fit to the cell model. The first inflection point of $|\bar{P}(r)|$
    is marked with a $\Box$.  For detailed parameters see system $1$ in
    Table \ref{basisdata}}
  \label{effcharge}
\end{figure}

If the counterions are monovalent, and we are dealing with a salt-free
solution, then $|\bar{P}(r)|$ also has the meaning of an integrated
probability distribution function for finding an ion at distance $r$.  In
Fig.~\ref{effcharge} we have plotted $|\bar{P}(r)|$ together with
$\barqeff(r)$ as a function of $\log (r)$ (see System 1 in Table.
\ref{basisdata}).  $|\bar{P}(r)|$ has two inflection points, the first is in
the inner region near the polyelectrolyte ($r<R_g$) the second further out
($r>R_g$).  Because only the inner region has an approximately cylindrical
symmetry the outer region cannot be compared to the cell model.  For
comparison we have plotted the integrated distribution function obtained from
the cell model for an infinitely long charged rod at the same concentration
with a value of $\xi_{CM}=1.65$.  This value was determined from the location
of the first inflection point of the simulated $|\bar{P}(r)|$, ($R_M$,
$|\bar{P}(R_M)|$), where the so-called Manning radius $R_M$ is the $r$ value
of the inflection point. Our simulation data yielded ($6.9\sigma \pm
1.0\sigma$,$0.39 \pm 0.03$), and leads to a charge parameter via $\xi_{CM} =
\frac{1}{1-|\bar{P}(r)|}$.  The other two parameters of the cell model $r_0$
and $R_C$ are fixed through the particle size $\sigma$ and the concentration.
The good agreement of the cell model curve with our simulated data for $r <
R_g$ suggests that the overall ion distribution is mainly governed by the
central part of the polyelectrolyte where we have a cylindrical symmetry in
the vicinity of the chain.  Also the $r$ value of the inflection point of the
fitted curve $R_M=6.67\sigma$, the so called Manning radius, is in good
agreement with the simulated data even though this value is fixed and not
fitted.  For $r>R_g$ the difference to the cell model is due to the loss of
the cylindrical symmetry in the simulated solution. This is especially
reflected by the occurrence of a second inflection point for $|\bar{P}(r)|$
which is not present in the solution of the cell model for the salt-free case.

However, we want to remark that the results for $|\bar{P}(r)|$ are neither
consistent with what one would expect inside the cell model from the bare line
charge parameter $\xi=0.98$ nor an effective line charge parameter
$\xi_{R_e}=\ell_B N_q / R_e=2.4$ determined by the chain extension.  This is
probably due to the complex geometrical situation around a flexible polymer.
From far away the polyelectrolyte can be seen as a rodlike object of length
$R_e$ with an effective line charge parameter $\xi_{R_e}$ but for closer
distances the local chain structure is important with an upper bound of $\xi$
for the line charge parameter characteristic for that region. Thus it is not
surprising that the fitted simulation result $\xi_{CM}$ is in between the two
extreme values of the line charge parameter.  We believe that this is due to
several differences between the simulated system and the cell model. First the
length of the polyelectrolyte is of the same order than the cell radius, thus
chain end-effects will play an important role. Second the configurations of
the polyelectrolyte can only roughly be approximated by a cylindrical stiff
object.  And last the solution of the cell model on the level of the
Poisson-Boltzmann equation neglects correlations between the particles.  The
inhomogeneities of the ion distribution along the polyelectrolytes backbone
which we will discuss from now on in this paper can not be compared to the
cell model since they are excluded from this theory by definition. A more
detailed analysis of the applicability of the cell model to flexible
polyelectrolytes will be left for future investigations.

\subsection{Qualitative Comparison with Scaling Theory}

As already mentioned above, a strongly charged polyelectrolyte with a fixed
charge distribution with mobile counterions can be looked upon like a
titrating polyelectrolyte with an annealed charge distribution where the
counterions together with their ability to neutralize a fraction of the chain
charges, play the ``annealed'' part.  We will demonstrate that our strongly
charged polyelectrolytes regarding the inhomogeneity of the effective charge
behaves qualitatively the same way as scaling theory predicts for titrating
polyelectrolytes\cite{castelnovo00a}.  We will also show in the following
subsections that the dependence of this inhomogeneity from system parameters
are in good agreement.  One of the observable effects is the accumulation of
counterions in the middle parts of the chain, as can be inspected in Fig.
{\ref{qeff_rc}}. Towards the end of the chain, the effective charge increases
which is equivalent to a decrease in the counterion concentration.  The reason
for this behavior is the difference in the electrostatic potential created by
the charges of the polyelectrolyte, which is stronger in the middle than at
the ends. In a simplified picture one can say that an ion close to the middle
part of the chain is attracted by more chain charges than an ion sitting at
the chain end.  This effect as well as the electrostatic field of a charged
flexible chain itself has been analytically described by Castelnovo et al. for
titrating polyelectrolytes \cite{castelnovo00a}.

Apart from this common origin of the inhomogeneity and the similarity of the
results there are some important differences between the two systems which
restricts the comparison.  The strength of the electrostatic interaction in
the theory and in the simulation is quite different. Thus, the blob picture
used in the scaling theory, is not applicable for our simulations.  In the
scaling theory of the titrating polyelectrolytes the charge fraction $f$ fixes
both, the electrostatic field of the chain and the number of charges subject
to this field.  This is different from our system, where $f$ fixes the
quenched charge distribution, and thereby the electrostatic field of the
chain, but the number of considered charges subject to this field is
determined by the number of ions inside the cut-off distance $r_c$.  At first
sight there is also a difference in the origin of the charge inhomogeneity.
In the case of a titrating polyelectrolyte, the inhomogeneity is due to a
repulsion between the mobile charges on the chain. In our system, the
inhomogeneity is due to the attraction of the free counterions to the fixed
chain charges.  When, however, the free ions close to the chain and the chain
charges are considered together as an ``effective charge'', then it is again
the repulsion between the effective charges of the chain which causes the
inhomogeneity.

The comparison between scaling theory and our simulations can only be
qualitative because the compared systems are physically different.  Thus
qualitative means that the predictions of the scaling theory about the
end-effect and its parameter dependencies can be confirmed.  Although the
functional form of the end-effect is very similar between the two systems we
cannot perform a quantitative comparison, e.g. fitting the results of the
scaling theory to our data, due to the different meaning of the charge
fraction $f$ of the scaling theory and our effective charges.

The result of charge accumulation at the end is also in agreement with results
from simulational studies of both flexible and rigid weakly charged titrating
polyelectrolytes where the electrostatic interactions are treated on the
Debye-H\"uckel level \cite{ullner96a,berghold97a}.

\subsection{Chain length}

\begin{figure}[htb]
  \begin{center}   
    \input{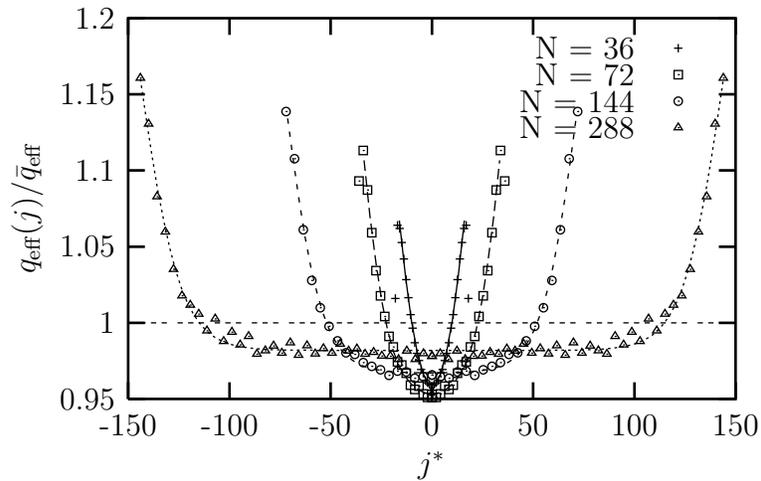}
  \end{center}
  \caption{Effective charge along the contour length
    $\qeff(j)$ for different chain length $N_m$. Systems
    2, 3, 4, 5 with increasing length. ${j}^*$: Here we have shifted the
    monomer index $j$ such that the innermost monomer has $j=0$. }
  \label{qeff_ndep}
\end{figure}

Because of the already mentioned influence of the surrounding geometry it is
first of all interesting to look at the chain length dependence of the
end-effect.  In Fig. \ref{qeff_ndep} the inhomogeneity of the effective charge
is shown for different chain length $N_m$, varying from $N_m = 36$ to $N_m =
288$ (systems 2,3,4, and 5 in Table \ref{basisdata}).  The qualitative form,
the accumulation of charge at the ends of the chain, does not change with
chain length. However, one observes a slight increase of the amplitude of the
end-effect from $0.11$ over $0.16$, $0.17$ to $0.18$ with increasing chain
length.  For the longest chains with $N_m \ge 144$ one can see a plateau,
which indicates a finite penetration depth of the end-effect.  The amplitude
depends on the difference of the electrostatic field of the chain and on the
number of annealing ions with distance $r$ smaller than $r_c$. The slight
increase of the amplitude points to a saturation of both quantities already
for chains with length $N_m \ge 72$.  An estimation for the case $N_m = 288$
gives a penetration depth of $65$ monomers along the contour length which
corresponds to a length of approximately $45 \sigma$ in real space. This
penetration depth is in agreement with the extension of the other chains. One
has to compare the penetration depth with $R_e/2$ which is $11.5\sigma$,
$24\sigma$, $49.5\sigma$ and $97.5\sigma$ for our chains with increasing
length. This explains why there is no plateau visible for the shorter chains.
Our observed penetration depth of $45\sigma$ fits well in magnitude to the
Debye-length $\lambda_{D} = 39.9 \sigma$, which was calculated assuming the
same ion concentration and an isotropic electrolyte solution.  However, our
limited precision does not allow us to study further corrections to the
screening length due to the inhomogeneity of our systems.

We are aware that the system with $N=288$ is on the edge to the semi-dilute
regime where the chains counterion clouds start to overlap but the influence
on the chain extension and the local structure is so small that it does not
influence our results significantly. The simulation time for the longest
chains was roughly 8 CPU days on a 600MHz workstation with a Compaq EV6 chip.
Due to limited CPU time and thus large statistical errors for long chains we
use shorter chains for the investigation of the other parameters.

\subsection{Charge parameter $\xi$}

\begin{figure}[htb]
  \begin{center}   
    \input{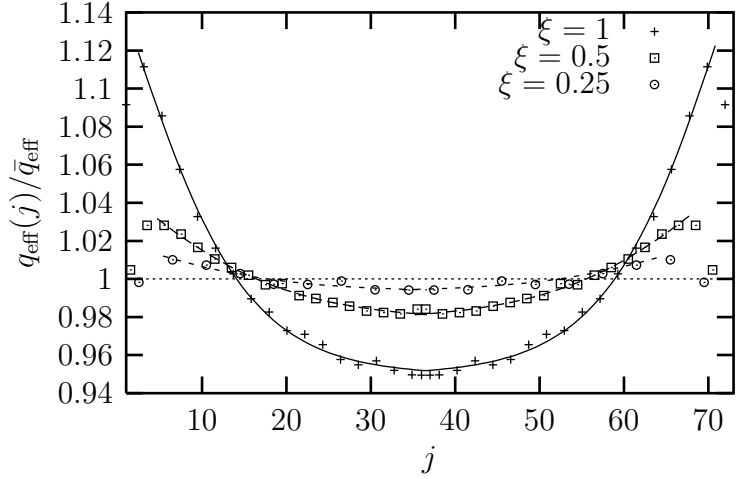}
  \end{center}
  \caption{Effective charge along the contour length
    $\qeff(j)$ for
    different values of the charge parameter $\xi$. Systems 3, 6, 7 with
    decreasing $\xi$. }
  \label{qeff_xidep}
\end{figure}

In this section we vary $\xi$ from $1$ over $0.5$ to $0.25$ by changing the
charge fraction $f$ and keeping the chain length $N_m=72$, $71$ and $69$
respectively roughly constant (Systems 3, 6, and 7 in Table~\ref{basisdata}).
Lowering $\xi$ yields a dramatic decrease of the end-effect as can be seen in
Fig. \ref{qeff_xidep}. The amplitude diminishes drastically from $0.16$ over
$0.044$ to $0.016$ with decreasing $\xi$. This result is again in good
qualitative agreement with results from scaling theory (compare Fig.2 in Ref.
\cite{castelnovo00a}).  One reason of the observed behavior can be traced to
the diminished electrostatic potential difference of the chain along its
contour that allows the counterions to distribute more and more uniformly
along the chain.  Not only the difference responsible for the end-effect is
getting smaller but also the electrostatic field of the chain itself becomes
weaker, thereby allowing the counterions to move further away from the chain.
This in turn yields fewer counterions which can play the annealed part, and
also results in a reduced amplitude of the end-effect.  We therefore conclude
that the concept of an annealed charge distribution mediated by the free ions
is not applicable to weakly charged chains with $\xi \ll 1$.  The increase of
the effective charge $\barqeff$ from $0.83e$ over $0.93e$ to $0.96e$ which
corresponds almost to the bare charge of the chain underlines our argument.

As is well known the charge parameter influences also the chain conformation.
With decreasing $\xi$ (and less repulsion between the chain charges) the chain
shrinks from $R_e=48\sigma$ over $R_e=34\sigma$ to $R_e=23\sigma$.  This slows
also down the decrease of the effective charge parameter $\xi_{R_e} = \ell_B
N_q / R_e$, which is lowered from $1.5$ over $1.0$ to $0.75$, and thus is only
half as large for the first system, whereas the bare $\xi$ has been reduced to
a quarter. However, it is for all systems still close to the critical value
$1$. As we have described in section \ref{cellmodel}, $\xi_{R_e}$ gives a
better estimate for the counterion attraction at large distances than $\xi$
itself.  But the reduced decrease of $\xi_{R_e}$ can not prevent the loss of
counterions into the bulk solution ,which again gives less possibilities of
annealing, and results in a large variation of the end-effect.

\subsection{Polymer concentration and added salt}

\begin{figure}[htb]
  \begin{center}   
    \input{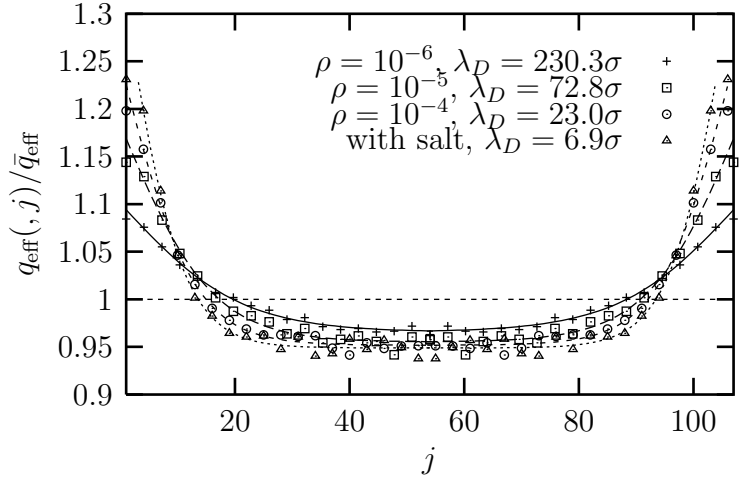}
  \end{center}
  \caption{Effective charge along the contour length
    $\qeff(j)$ (here with a cutoff radius
    $r_c=3.14\sigma$) for different densities and salt concentration. Systems
    1 ($\rho=10^{-4}$), 8 ($\rho=10^{-5}$), 9 ($\rho=10^{-6}$) and system 10
    ($\rho=6 \cdot 10^{-4}$) with salt.}
  \label{qeff_kdep}
\end{figure}

In this section we investigate the influence of the polymer concentration and
added salt on the end-effect. The results for systems $1$, $8$, $9$ and $10$
are shown in Fig \ref{qeff_kdep}.  With decreasing density the amplitude of
the end-effect is decreasing from $0.25$ for $\rho=10^{-4}\sigma^{-3}$ over
$0.19$ for $\rho=10^{-5}\sigma^{-3}$ to $0.12$ for $\rho=10^{-6}\sigma^{-3}$.
Also the penetration depth is decreasing with increasing concentration. This
can be seen from the formation of a plateau region for the higher densities.
The decrease of the amplitude is due to the effect that with decreasing
polymer concentration the counterions will explore the larger accessible
volume, or more technical spoken, they will gain more translatorial entropy,
and thus more counterions will be found at larger distances. Thus less
counterions can feel the inhomogeneity of the electrostatic field of the chain
which makes the end-effect less pronounced. This is reflected in an increase
of $\barqeff$ upon dilution from $\barqeff=0.65e$ for
$\rho=10^{-4}\sigma^{-3}$ to $\barqeff=0.88e$ for $\rho=10^{-6}\sigma^{-3}$.
But also the decrease of $\xi_{R_e}$ from $2.4$ at $\rho=10^{-4}\sigma^{-3}$
to $1.9$ at $\rho=10^{-6}\sigma^{-3}$ coming from the slight increase of the
chains extension adds to the decrease of the amplitude. The decrease of the
penetration depth with increasing density can be explained by the better
screening at higher ionic strength. We have chosen a smaller value for the
cutoff $r_c=3.01\sigma$ for the data in Fig. \ref{qeff_kdep} in order to have
a smaller cutoff radius than one Debye-length.  An addition of salt to system
$1$ keeps the polyelectrolyte density fixed, but enhances the charge density
further to $\rho=6 \cdot 10^{-4}\sigma^{-3}$, and we observe the same trends
as before, namely that the amplitude is increased from $0.25$ to $0.29$.

The results depicted in Fig.  \ref{qeff_kdep} might at first sight appear
puzzling, because for a smaller Debye-length one would expect a decrease of
the amplitude of the end-effect due to the fact that then each ion can 'see'
only a smaller part of the chain, thus making the difference between a
location at the middle and the end of a chain weaker.  The reason for this
somewhat unexpected behavior is that the dominating factor for the amplitude
of the end-effect is not the change of the screening length $\lambda_D$ but
the change in the number of annealing ions. 

\begin{figure}[htb]
  \begin{center}   
    \input{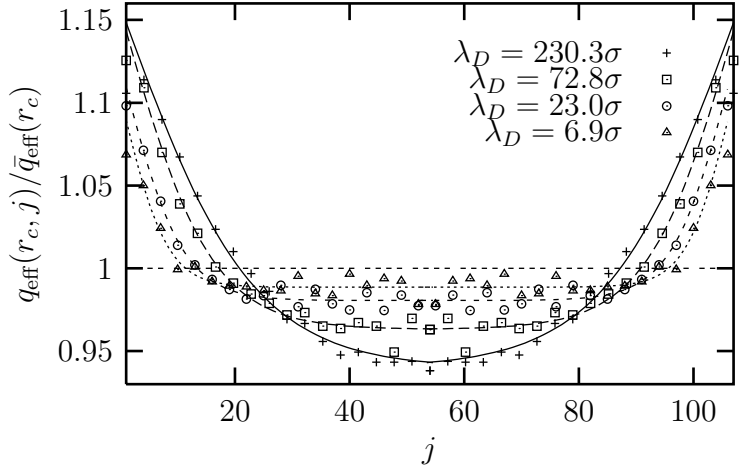}
  \end{center}
  \caption{Effective charge along the contour length
    $\qeff(j)$ (here with constant
    $\barqeff=0.85$) for different densities and salt
    concentration as in Fig. \ref{qeff_kdep} }
  \label{qeff_const}
\end{figure}

To investigate the influence of the screening length and to be able to test
the predictions of the scaling theory we have done the data analysis also for
constant $\barqeff$ instead of constant cutoff radius $r_c$.  The scaling
theory predicts for this case a decrease of the amplitude of the end-effect
with decreasing screening length (compare Fig 3. in Ref.
\cite{castelnovo00a}).  This can also be confirmed by our data, compare Fig.
\ref{qeff_const}, where all systems have the same average effective charge
$\barqeff=0.85e$ and thus different values for $r_c$ ranging form $1.5\sigma$
to $14\sigma$. Here the amplitude of the end-effect changes from $0.18$ at
$\rho=10^{-6}\sigma^{-3}$ to $0.16$ at $\rho=10^{-5}\sigma^{-3}$ to $0.12$ at
$\rho=10^{-4}\sigma^{-3}$ and $0.09$ for the salt case which is the predicted
behavior. The changes in the penetration depth can also be observed more
clearly in Fig.~\ref{qeff_const}, and yield the same values as those derived
from Fig.  \ref{qeff_kdep}.

\subsection{Effects on the conformation}

\begin{figure}[htb]
  \begin{center}   
    \input{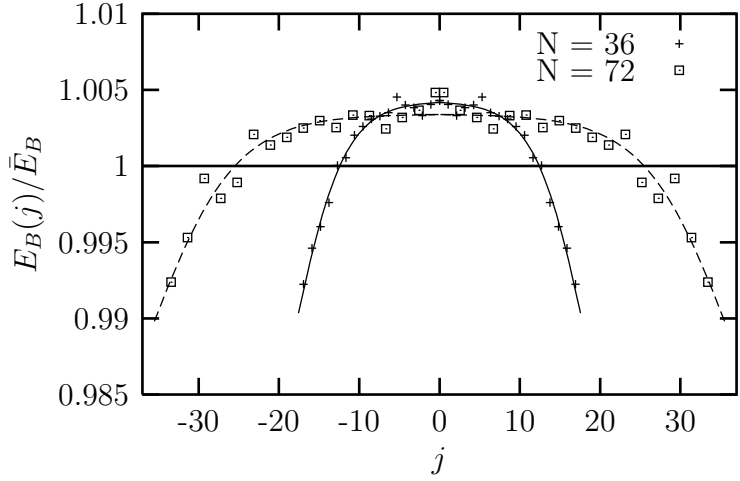}
 \end{center}
  \caption{Configurational inhomogeneity: Relative bond energy
    $E_B(j)/\bar{E}_B$ along contour length $j$ for different chain length
    $N_m$. Systems 2 and 3.}
  \label{conf}
\end{figure}

Besides the inhomogeneity of the local effective charge there is also an
inhomogeneity in the local conformation of a strongly charged polyelectrolyte.
This inhomogeneity has different origins.  Here we are interested in effects
caused by the electrostatic interaction of the charged particles.  The same
inhomogeneity in the electric field that causes the free ions to distribute
inhomogeneously also acts on the fixed charges of the polyelectrolyte. This
gives rise to different stretching of the bonds along the contour length. One
can see this effect by measuring the energy stored in the bonds. The relative
bond energy $E_B(j)/\bar{E}_B$ is shown in Fig. \ref{conf} for two different
chain length (Systems 2 and 3 in Table \ref{basisdata}). Here $E_B(j)$ is the
energy stored in the bond between the monomers $j$ and $j+1$ coming from the
Lennard-Jones and FENE potentials as defined in equations \ref{ljpotphob} and
\ref{fenepot} and $\bar{E}_B$ is the average of $E_B(j)$ for the whole chain.

From a theoretical point of view it is interesting to know something about the
tension along the chain, since the chain under tension model is the basis for
the electrostatic blob picture used in scaling theory.  Thus we have measured
the bond energy along the contour length of the chain.  The energy and thus
the tension is lower at the ends of the chain. This yields in the blob picture
a bigger electrostatic blob size at the ends than in the middle. This causes
the polyelectrolyte to appear in a trumpet like shape \cite{castelnovo00a}.

There are at least two reasons why we are not able to see this effect in a
more direct fashion. The first one is that with our strongly charged chains we
are far away from the region were the blob picture is applicable. Since for
strongly charged chains an electrostatic blob would contain roughly only one
monomer we can not measure the blob size directly via, e.g., the scaling of
local distances.  It is also impossible to relate the transversal extension of
the chain perpendicular to its main axis (defined for instance by the
principle axis of inertia) to the inhomogeneity of the electrostatic field of
the chain. This is because already for neutral chains it is known, that the
transversal fluctuations have different magnitudes along the chain, namely
they are larger at the ends of the chain than in the middle \cite{kantor99a}.

\subsection{Poor solvent condition}

\begin{figure}[htb] 
  \begin{center}
    \epsfxsize=8.5cm \epsfbox {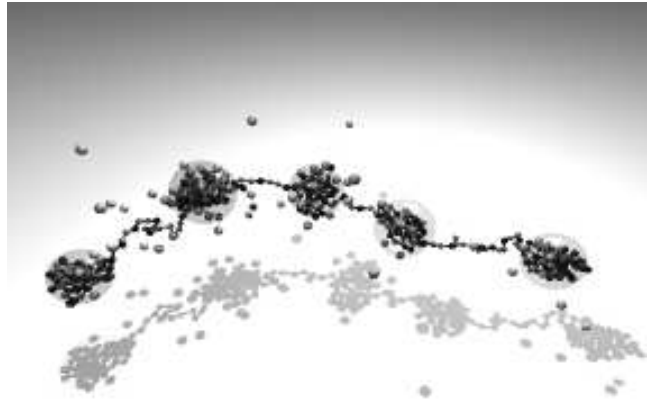}
  \end{center}
  \caption{Typical necklace conformation. Chain with charged (black) and
    neutral (grey) monomers, pearl extension (transparent grey), counterions
    (light grey). System 11.}
  \label{pn_snap}
\end{figure}

\begin{figure}[htb]
  \begin{center}
    \input{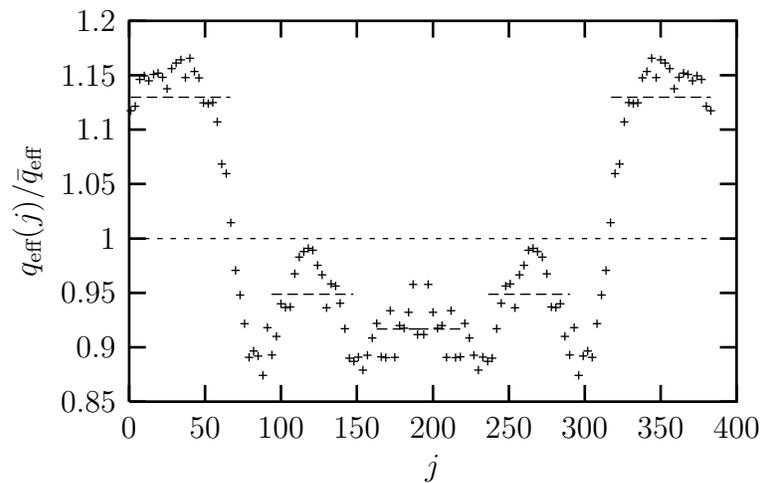}
  \end{center}
  \caption{Effective charge along the contour length
    $\qeff(j)$ for the poor
    solvent case (pearl necklace structure). Pluses and solid line shows the
    data. The dotted horizontal lines indicate the location and the effective
    charge of the substructures (only pearls). System 11.}
  \label{qeff_pn}
\end{figure}

In the case of polyelectrolytes under poor solvent conditions the additional
short range attraction between the backbone monomers of the chains makes the
already impressing variety of phenomena known from the good solvent case even
richer and more interesting. Due to the balance of attractive and repulsive
forces so-called necklace structures occur
\cite{dobrynin96a,dobrynin99a,lyulin99a,micka99a,micka00a}.  Here we have
simulated a system with $\epsilon_{LJ}=1.75$ which is deep in the poor solvent
regime \cite{micka99a,micka00a}.  For the system (No. 11 in Table
\ref{basisdata}) under investigation we find an interesting coexistence
between structures with four and five pearls.  Fig.  \ref{pn_snap} shows a
snapshot of a necklace with five pearls. The question of the stability of
these structures for strongly charged systems will be investigated in a future
publication. Here we concentrate only on the difference in the counterion
distribution around the strings and pearls.  In order to see the relation
between a particular necklace conformation and the local effective charge we
restrict the data analysis to configurations with five pearls. The number of pearls
in a configuration is calculated via a cluster algorithm which is adapted for
linear chains \cite{limbach01b}. It is clearly visible in Fig. \ref{qeff_pn}
that the end-effect in this case is smeared out almost uniformly over the end
pearls. There is a sharp decrease in the effective charge in the region of the
first string.  In this gap between two pearls we find an accumulation of
counterions and thus an enhanced decrease of the effective charge.  The
necklace structure in the middle part of the chain is also visible in the
modulation of the effective charge along the backbone. Fig. \ref{qeff_pn}
shows the effective charge along the contour length together with the
effective charge of the five pearls.

However, with our method of measuring the effective charge it is not possible
to answer the question whether the strings or the pearls in such a structure
have different effective charges as it is proposed in \cite{castelnovo00a}.
The difficulty is due to the limited size of the strings, and the difficulty
to decide for the counterions between two pearls, if they belong to one of the
pearls or to the string between them.

%
%
\section{Conclusion}
%
%
In this paper we looked at the spatial distribution of the counterions around
strongly charged polyelectrolytes by means of MD simulations. We demonstrated
that by partially neutralizing the quenched charged distribution on the chain
backbone the inhomogeneous distribution of counterions lead to the same
qualitative effects that are observed in weakly charged polyelectrolytes with
an annealed charge distribution.

We have discussed the difficulty in defining the effective charge through the
cut-off radius, because this radius also fixes the average charge fraction,
but have also shown that the qualitative appearance of the end-effect is
rather independent from its definition.  The comparison of the simulated ion
distribution with the Poisson-Boltzmann solution of the cell model of an
infinitely extended charged rod has shown that the description of our
polyelectrolytes as rodlike objects is valid which allows a description of the
inhomogeneities along the contour of the chain in the presented way.  We made
a qualitative comparison with the results obtained for weakly charged
titrating polyelectrolytes via scaling theory by Castelnovo et al.
\cite{castelnovo00a}.  We discussed the common underlying physical mechanism,
namely the differences in the electrostatic field of the chain along its
backbone, and showed that this assumption is compatible with our results.  The
agreement between predictions of the scaling theory and our simulations was
also confirmed for the qualitative dependencies of the end-effect from the
investigated parameters chain length, charge fraction and ionic strength.  We
found a saturation of the end-effect for long chains, when the chain
extension, namely $R_e$, is at least twice as large as the Debye screening
length.  A simple Debye-length criterion appears to be sufficient to explain
the penetration depth of the end-effect.  However, when we looked at the
amplitude dependency on density and ionic strength of the solution, we found
that in this case both parameters, the number of annealing ions and the ionic
strength of the solution, influence the end-effect and that the first one
dominates.  We therefore fixed the number of annealing counterions to
investigate the ionic strength dependence. This enabled us also to observe the
expected decrease of the end-effect with increasing ionic strength.  The
amplitude of the end-effect was shown to depend strongly on the charge
parameter $\xi$. The definition of such an end-effect via close mobile
counterions can not be made for an effective charge $\xi << 1$.  We observed
the same conformational inhomogeneities that are found for weakly charged
annealed polyelectrolytes.

Even though the chain conformation is very different in the poor solvent case
the end-effect is qualitatively the same, namely the counterions are more
likely to be found at the middle of the chain than at the ends.  We could also
clearly see the necklace structure by looking at the effective charge along
the contour length. However the string length was to short to show any charge
difference in pearls and strings as has been predicted in Ref.
\cite{castelnovo00a} Overall we can conclude that the charge distribution of
strongly charged polyelectrolytes (with or without annealing) behaves like
that one of weakly charged titrating polyelectrolytes. This is due to the
presence of the mobile partially neutralizing counterions, which results in an
annealed backbone charge distribution.  We hope that this work stimulates the
future development of theories for strong polyelectrolyte solutions as well as
for inhomogeneous electrostatic systems in general.

\section{Acknowledgments}
We thank M. Castelnovo and J.-F. Joanny for helpful and clarifying remarks
concerning their scaling theory, and acknowledge many discussions within the
theory group at the MPI, especially M. Deserno, K. Kremer, and R. Messina.  A
large computer time grant hkf06 from NIC J\"ulich and financial support by the
German Science foundation is also gratefully acknowledged.


\begin{table}[htbp]
\begin{tabular}{|c|c|c|c|c|c|c|c|c|c|c|} \hline
  System & $N_p$ & $N_m$ & $N_q$ & $\rho$ & $R_e$ & $R_g$ & $\ell_B$ & $\xi$ &
  $\xi_{R_e}$ & $\barqeff$\\[.5ex] \hline
  1  & 8  & 106 & 36  & $10^{-4}$ & 46 & 16 & 3.0 & 0.98 & 2.4 & 0.65\\ \hline
  2  & 7  & 36  & 36  & $10^{-4}$ & 23 &  8 & 1.0 & 0.98 & 1.6 & 0.87\\ \hline
  3  & 5  & 72  & 72  & $10^{-4}$ & 48 & 16 & 1.0 & 0.98 & 1.5 & 0.83\\ \hline
  4  & 3  & 144 & 144 & $10^{-4}$ & 99 & 31 & 1.0 & 0.98 & 1.4 & 0.80\\ \hline
  5  & 3  & 288 & 288 & $10^{-4}$ &195 & 62 & 1.0 & 0.98 & 1.5 & 0.79\\ \hline
  6  & 5  & 71  & 36  & $10^{-4}$ & 34 & 12 & 1.0 & 0.49 & 1.0 & 0.93\\ \hline
  7  & 5  & 69  & 18  & $10^{-4}$ & 23 & 8  & 1.0 & 0.25 &0.75 & 0.96\\ \hline
  8  & 8  & 106 & 36  & $10^{-5}$ & 52 & 17 & 3.0 & 0.98 & 2.1 & 0.77\\ \hline
  9  & 8  & 106 & 36  & $10^{-6}$ & 56 & 18 & 3.0 & 0.98 & 1.9 & 0.88\\ \hline
  10 & 4  & 106 & 36  &$6\cdot 10^{-4}$ & 35 & 13 & 3.0 & 0.98 & 3.1 & 0.42\\ \hline
  11 & 5  & 382 & 128 & $10^{-5}$ & 46 & 17 & 1.5 & 0.49 & 4.2 & 0.59\\
\hline
\end{tabular}
\vspace{2ex}
  \caption{Some basic observables for the used systems:  $N_p$ Number of
    polymers in simulation box, $N_m$ chain length (number of monomers), $N_q$
    charge of polyelectrolyte, $\rho$ charged particle density  in
    $\sigma^{-3}$ (number of charged particles divided by the box volume),
    $R_e$ end-to-end distance, $R_g$ radius of gyration, $\xi$ charge
    parameter along the contour (average bond length is $1.02\sigma$), $\xi_{R_e}$ charge parameter regarding $R_e$,
    $\barqeff(r_c=5.01)$ effective charge in $[e]$ of a
    monomer and its ions closer than $r_c$. All length are given in units of
    $\sigma$. The statistical error of the measured quantities is less than 3\%.}
  \label{basisdata}
\end{table}
\clearpage

%
%
%
%
%
%
\clearpage

\clearpage

\clearpage

\clearpage

\clearpage

\end{document}